# Study of plasmon resonance in a gold nanorod with an LC circuit model

Cheng-ping Huang[1,2*], Xiao-gang Yin[1], Huang Huang[1], and Yong-yuan Zhu[1*]

[1]*National Laboratory of Solid State Microstructures, Nanjing University*

*Nanjing 210093, P.R. China*

[2]*Department of Applied Physics, Nanjing University of Technology*

*Nanjing 210009, P.R. China*

Abstract

Gold nanorod has generated great research interest due to its tunable longitudinal plasmon resonance. However, little progress has been made in the understanding of the effect. A major reason is that, except for metallic spheres and ellipsoids, the interaction between light and nanoparticles is generally insoluble. In this paper, a new scheme has been proposed to study the plasmon resonance of gold nanorod, in which the nanorod is modeled as an LC circuit with an inductance and a capacitance. The obtained resonance wavelength is dependent on not only aspect ratio but also rod radius, suggesting the importance of self-inductance and the breakdown of linear scaling. Moreover, the cross sections for light scattering and absorption have been deduced analytically, giving rise to a Lorentzian line-shape for the extinction spectrum. The result provides us with new insight into the phenomenon.

\* Email: cphuang@njut.edu.cn, yyzhu@nju.edu.cn

## 1. Introduction

It is well known that the metallic particles will undergo the plasmon oscillations when illuminated by an incident light, where the confined conduction electrons can be driven by the electric field into resonance [1]. In addition to maxima of light scattering and absorption, the plasmon resonance is accompanied by a strong field enhancement inside and around metallic particles. Such effect has generated great interest due to its wide range of potential applications in, such as, signal amplification, molecular recognition [2], as well as cancer diagnosis and therapy [3]. Moreover, since the size of metallic particle is much smaller than the resonance wavelength, this effect enables us to manipulate light with a wavelength of microns by using nanoscale objects. For example, employing a plasmonic waveguide consisting of closely spaced nanoparticles, light propagation below the diffraction limit can be achieved [4].

Recent advances in nanotechnology have lead to successful synthesis of non-spherical nanoparticles such as gold nanorods [5, 6], which typically has a length of 40~200nm and diameter of 10~30nm. Compared with a spherical metallic particle, a gold nanorod has a tunable longitudinal plasmon resonance, which is particularly desirable for practical applications [3, 7]. To deeply exploit the resonance effect, a full understanding of the interaction between light and nanorods is urgently needed. Conventionally, the interaction of light with metallic particles is investigated by solving the Laplace or Maxwell equations and matching the fields at the metal surface [8]. This works well for the metallic spheres and ellipsoids but is not suitable for nanorods. Numerical simulation using various methods may provide another solution for this problem [9-11], but unable to reveal the underlying physics. Currently, a widely used method is to employ the Gans theory, which was developed more than ninety years ago for spheroidal particles, to approximately describe the plasmon resonance of gold nanorod [6, 12, 13]. According to the Gans theory, the resonance wavelength is only a function of aspect ratio and medium permittivity, and in certain conditions a linear relationship between them can be resulted [12]. Nonetheless, recent numerical results suggest that, even when the aspect ratio is fixed and the retardation effect is weak, the position of longitudinal resonance can still vary strongly with the rod radius [9, 10]. This indicates that the detailed physics for the plasmon resonance of gold nanorod has not been well captured and understood.

Here, we will break through the confines of Gans theory and propose a new scheme to study the longitudinal plasmon resonance of a gold nanorod. In this scheme, the problem is treated simply using an LC circuit model without solving the Laplace or Maxwell equations. The obtained resonance wavelength is dependent on not only aspect ratio but also rod radius, in agreement with the numerical simulations. The result shows that the radius dependence is related to the self-inductance and that a breakdown of linear scaling is also present due to the formal inductance associated with the inertia of electrons. Moreover, the cross sections for light scattering and absorption have been deduced analytically, giving rise to a Lorentzian line-shape for the extinction spectrum. The result overcomes the deficiency of Gans theory and provides us with new insight into the phenomenon.

**2. LC circuit model**

The structure under study is composed of a subwavelength gold nanorod, which has a length of $l$ and radius of $r_0$ ($r_0 < l$) and is embedded in a dielectric with the permittivity of $\varepsilon_d$. The incident light propagates with the electric field along the rod axis, thus exciting the longitudinal plasmon resonance (see figure 1a). Supposing that the radius of nanorod is smaller than the skin depth ($r_0 < \delta \sim 20nm$), the retardation effect can be neglected and the fields inside the nanorod can be taken to be homogeneous. We also assume that the size of nanorod is much larger than the Fermi wavelength ($r_0 \gg \lambda_F \sim 0.5nm$), so that the quantum effect can be ignored and a classic description of the effect is applicable. These assumptions can be satisfied considering the actual sizes of the nanorods [6].

We will show in the following that a gold nanorod can be treated as an LC circuit having an inductance and a capacitance. As we know, under the action of the incident light, a current flow $I$ will be generated in the nanorod, which is accompanied by a magnetic field with the amplitude proportional to the current. The self-inductance $L$ is induced by the magnetic field energy $w$ that is stored both inside and outside the conductor with $w = LI^2/2$ [14]. Since the internal part of the magnetic energy is much smaller than the external part, we can neglect the former in the calculations. A plot of the magnetic field distribution (see figure 1b) shows that the external magnetic energy is concentrated mainly near the nanorod, namely, in a cylindrical region with the length $l$, the inner radius $r_0$ and the outer radius $l/2$. By calculating the energy in this cylindrical region, the self-inductance is obtained as

$$L = (\mu_0 l / 2\pi) \ln(l/2r_0), \tag{1}$$

which is mainly dominated by the length of nanorod.

It is also significant that a formal inductance can be provided by the ac resistance of a nanorod. Under the driving of a time-harmonic field with the angular frequency $\omega$, the velocity of electrons and the current density cannot be in phase with the driving field due to the presence of electron inertia. As a result, a frequency-dependent and imaginary part will present in the ac resistance $R = R_0 - i\omega L_0$, where $R_0 = l/\sigma_0 s$ ($\sigma_0 = ne^2\tau/m$ is the dc conductivity, $s = \pi r_0^2$ is the area of cross section) and $L_0 = \mu_0 l / k_p^2 s$ ($k_p = \omega_p/c$, $\omega_p = \sqrt{ne^2/m\varepsilon_0}$ is the bulk plasma frequency, and $c$ is the light velocity). Therefore, an ac resistance will behave as a cascade of a dc resistance $R_0$ and a formal inductance $L_0$. We note that the formal inductance can also be derived by calculating the electronic kinetic energy $w_0 = Nmv^2/2$ and setting $w_0 = L_0 I^2 / 2$ [15], where $v = -I/nes$ is the electron velocity and $N = nsl$ is the total number of electrons. Since the formal inductance is inversely proportional to the cross section, it will play a crucial role in the nanocircuit.

As a consequence of the current flow, electric charges with different signs $\pm q$ will accumulate on the opposite ends of the nanorod (where a homogeneous distribution of charge on the end faces can be assumed). Correspondingly, the two end faces (circular disks) of the nanorod will function as one circular capacitor. According to the electrostatics theory that is applicable here, the potential of one charged circular disk can be determined theoretically and a potential difference between the two oppositely charged disks will be expected. It should be mentioned that the potential of a circular disk is not homogeneous but decreases gradually from the center to the edge. Nonetheless, at the center of one disk, the potential is simple and can be worked out analytically with $U = q/2\pi\varepsilon_0\varepsilon_d r_0$ (the weak contribution from the other disk is firstly neglected). If the capacitance C is defined simply as the electric charges divided by the potential difference between the two disk centers, we immediately have $C = \pi\varepsilon_0\varepsilon_d r_0$. Certainly, the effective capacitance will be modified by the effects such as the inhomogeneous distribution of the potential and the weak coupling between the two disks. To take account of these effects, we phenomenologically introduce a correction coefficient $\alpha$ to the effective capacitance

$$C = \alpha\pi\varepsilon_0\varepsilon_d r_0. \tag{2}$$

The expression given by equation (2) is different from that of a commonly used parallel-plate capacitor, which shows that the effective capacitance of nanorod is mainly governed by the rod radius and the permittivity of surrounding medium. Here, the correction coefficient can be determined by comparing the analytically and numerically obtained resonance peaks. For the gold nanorod, an appropriate value of $\alpha = 2.5$ has been found and will be used in the following.

**3. Results and discussions**

In the quasi-static approximation, the Ohm equation applied to a subwavelength LC circuit, which is subject to an incident light with a time-harmonic field, is written as $\varepsilon = I(R - i\omega L + i/\omega C)$, where $\varepsilon$ is the electromotive force (emf). Different from the split-ring resonators (SRRs) having a negative permeability [16, 17], here the emf is provided by light electric field rather than the magnetic field, with $\varepsilon = \int E \cdot dl = El$. As is seen above, the gold nanorod, which carries the positive and negative charges on the opposite ends, can be regarded as an electric dipole with its dipole moment $p = ql$. By using the Ohm's law and the relationship $I = dq/dt$, we have

$$p = \frac{\varepsilon_0 A}{\omega_0^2 - \omega^2 - i\eta\omega} E, \qquad (3)$$

where $A = \mu_0 c^2 l^2 /(L_0 + L)$, $\omega_0 = 1/\sqrt{(L_0 + L)C}$, and $\eta = R_0/(L_0 + L)$. Therefore, the gold nanorod is a typical Lorentz resonator in response to the light.

According to equation (3), the gold nanorod has a plasmon resonance frequency of $\omega_0 = 1/\sqrt{(L_0 + L)C}$. Correspondingly, the vacuum wavelength of plasmon resonance, which is of particular interest, can be determined to be

$$\lambda_0 = \pi n_d \sqrt{10\kappa (2\delta^2 + r_0^2 \ln \kappa)}, \qquad (4)$$

where $n_d$ is the refractive index of the surrounding medium, $\kappa = l/2r_0$ is the aspect ratio of nanorod, and $\delta = c/\omega_p$ is the skin depth of gold (~21.9nm). In contrast to previous result based on the Gans theory [12, 13], two features can be pointed out here. Firstly, although a linear dependence was believed, equation (4) presents a nonlinear relationship between the resonance wavelength and the aspect ratio as well as the medium permittivity. Secondly, the resonance wavelength is not only a function of aspect ratio but also a function of the rod radius, thus overcoming the deficiency of Gans theory [10]. Since the gold nanorod is thin and the phase retardation is not considered here, the latter feature is not due to the retardation effect as believed but rather due to the self-inductance of the nanorod. This is not difficult to understand: if we neglect the self-inductance $L$ (when the rod radius is very small), the resonance wavelength is then dependent on the aspect ratio alone ($L_0$ and $C$ is proportional to $l/r_0^2$ and $r_0$ respectively, thus $L_0 C \propto \kappa$). Note that when the rod radius is large enough (not studied here), the retardation effect will also play an important role. The above result suggests, on the other hand, that a scaling down of the rod size cannot reduce the resonance wavelength infinitely, in agreement with previous reports on SRRs [15, 18]. This means a breakdown of linear scaling is also present in the nanorod, originating from the formal inductance associated with the inertia of electrons (neglecting the formal inductance will lead to a linear scaling $\lambda_0 \propto r_0$).

To test the performance of the theoretical result, we have compared the resonance wavelengths obtained by equation (4) with that of the experiments. Figure 2(a) shows the dependence of the resonance wavelength on the aspect ratio, where the refractive index of the surrounding medium (water) is 1.33 and the radius of gold nanorod is fixed as 11nm. The solid squares represent the measured values (the nanorods have a particle size distribution) [10], and the solid line gives our analytical results for the single nanorods. One can see that the analytical calculations agree well with the experiments, concerning the spectral position of plasmon resonance as well as its dependence on the aspect ratio (a deviation presents when the aspect ratio becomes very small, where the capacitance of nanorod will be less well defined). It should be mentioned that the actual geometry of a fabricated nanorod will deviate slightly from the modeled structure, which yields a reduced plasmon resonance wavelength [10]. Nonetheless, compared with a single nanorod, the resonance wavelength of an ensemble of nanorods, which has a particle size distribution and an average aspect ratio, will be shifted to a longer wavelength [10]. Consequently, these two effects will be largely canceled by each other.

By fixing the aspect ratio, we have also studied the dependence of resonance wavelength on the rod radius. Figure 2(b) shows both the analytically determined (solid lines) and numerically calculated (solid symbols taken from Ref. [19]) results, where the aspect ratio is set as 3, 5 and 7 respectively (from the bottom to the top) and the rod diameter is increased gradually from 5nm to 30nm. The analytical and numerical calculations agree well with each other, which suggest that, even when the aspect ratio is fixed, the resonance wavelength will still increase with the rod radius significantly. For the three aspect ratios 3, 5, and 7, a wavelength shift of 83nm (712~795nm), 154nm (920~1074nm), and 217nm (1090~1307nm) has been obtained respectively, which corresponds to a relative shift of wavelength about 12%, 17%, and 20%. In contrast, the Gans theory predicts a constant resonance wavelength, which is, respectively, 670nm, 860nm, and 1060nm for the above aspect ratios [10]. It can be deduced from figure 2(b) that the resonance wavelength in the surrounding medium is more than 20 times larger than the rod width. Hence, the retardation effect can be neglected in the propagation direction of light (note that, in the present incident configuration, there is no phase retardation along the rod axis).

A significant consequence accompanying the plasmon resonance of nanorod is the greatly enhanced light scattering and absoption. These effects can be calculated by introducing the rod polarizability $\chi$, which is linked to the dipole moment of nanorod via $p = \varepsilon_0 \chi E$. The cross sections for light scattering and absorption can be thus expressed, respectively, as $C_{sca} = k_0^4 |\chi|^2 / 6\pi$ and $C_{abs} = k_0 n_d^{-1} \text{Im}[\chi]$, where $k_0$ is the wavevector in free space. With the use of equation (3), we have

$$C_{sca} = \frac{A^2}{6\pi c^4} \frac{\omega^4}{(\omega_0^2 - \omega^2)^2 + \eta^2 \omega^2},$$
$$C_{abs} = \frac{A\eta}{n_d c} \frac{\omega^2}{(\omega_0^2 - \omega^2)^2 + \eta^2 \omega^2}.$$
(5)

For small gold nanorods, it is easy to find that $A \approx \omega_p^2 V$ and $\eta \approx \gamma$, where $V$ is the volume of nanorod and $\gamma$ is the collision frequency of electrons (here we set $\gamma = 1.5 \times 10^{14}$ rad/s, which is larger than that of the bulk metal due to additional scattering from the nanoscale metal surface [19]). Therefore, the maximal cross sections for scattering ($C_{sca}^m = k_p^4 \omega_0^2 V^2 / 6\pi \gamma^2$) and absorption ($C_{abs}^m = k_p^2 c V / n_d \gamma$) are strongly dependent on the volume of nanorod. Generally, the absorption is dominant for the small particles, and with the increase of rod size the scattering will become more significant. This agrees with the metallic spheres or ellipsoids.

With equations (5), the extinction cross section ($C_{ext} = C_{sca} + C_{abs}$) of a nanorod can be obtained theoretically. Without loss of generality, figure 3(a) shows the calculated extinction spectrum (the open circles) for a gold nanorod, which has a length of 52.65nm and a radius of 8.1nm. The spectrum exhibits a Lorentzian line-shape with the linewidth about 42nm. As a comparison, figure 3(a) also presents the measured extinction spectrum for the same (single) nanorod (the solid circles, see Ref. [11]). Although a slight shift of resonance wavelength (~55nm) is presented in the spectra (due to a deviation of actual geometry from the model, as mentioned above), a numerical fit of experimental data suggests, indeed, a Lorentzian line-shape (the solid line). In addition, figure 3(b) has plotted the calculated extinction spectra for three gold nanorods with different rod sizes (the radius is fixed as 10nm, and the length is set as 60, 80, and 100nm, respectively). Besides the red-shift of plasmon resonance peak, an increase of extinction cross section with the rod length or particle volume is clearly demonstrated.

The above scheme can also be employed to study the plasmon resonance of a rectangular nanorod. In this case, the resonance wavelength is deduced to be

$$\lambda_0 = \pi n_d \sqrt{\frac{5l\,(2\pi\delta^2 + ab\ln\kappa)}{a\ln(b/a+\sqrt{1+b^2/a^2}) + b\ln(a/b+\sqrt{1+a^2/b^2})}}. \tag{6}$$

Here, $\kappa \approx (l/2)\sqrt{\pi/ab}$, $l$ is still the length of nanorod, $a$ and $b$ are the side lengths of the cross section. Equation (6) predicts a nonlinear relationship between the resonance wavelength and the rod side length. Generally, the larger the side length of cross section is, the smaller the resonance wavelength. We stress that, even when the rectangular nanorod is folded into a U-shaped SRR, the effective inductance and capacitance are just slightly modified (note that it is still the end faces but not the gap of the U-shaped SRR that forms the capacitor). Hence, the main characteristics of the plasmon resonance of rectangular nanorod can be well maintained. This explains why a U-shaped SRR and a cut-wire share the same resonance features and why the resonance frequency increases nonlinearly with the gold thickness [20].

## 4. Conclusions

In summary, the interaction of light with a single gold nanorod has been studied analytically. The employment of an LC circuit model greatly simplifies the problem and captures well the characteristic of the phenomenon. The results show that the gold nanorod behaves as a Lorentz resonator in response to light and that the extinction spectrum has a Lorentzian line-shape. The plasmon resonance wavelength relies on not only aspect ratio but also rod radius, thus overcoming the deficiency of Gans theory. In contrast to conventional wisdom, the effect is actually related to the self-inductance of nanorod rather than the phase retardation. It is worthy of noticing that, according to the theory, a breakdown of linear scaling is also present in the nanorod. Our result represents a new understanding of the phenomenon.

This work was supported by the State Key Program for Basic Research of China (Grant Nos. 2004CB619003 and 2006CB921804), by the National Natural Science Foundation of China (Grant Nos. 10523001, 10804051, and 10874079).

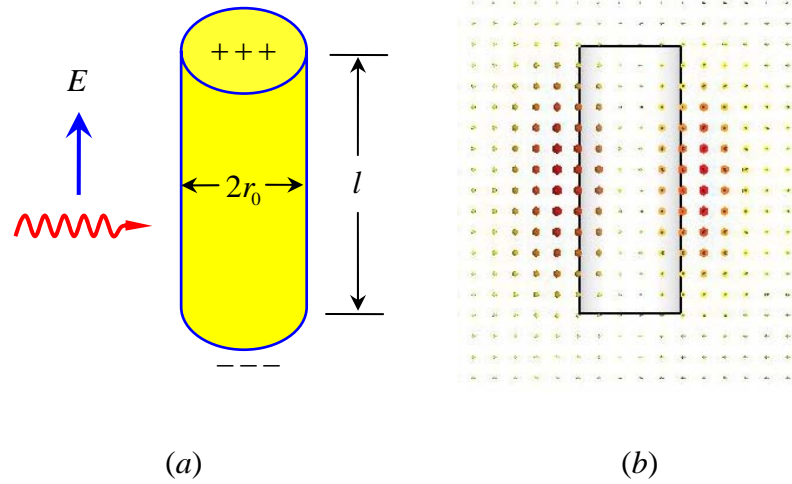

Figure 1: (a) Schematic view of the structure under study. The subwavelength gold nanorod is embedded in a dielectric, and the incident light is propagating with the electric field along the rod axis, thus exciting the longitudinal plasmon resonance. (b) The magnetic field distribution around the nanorod when a current flow in the rod is excited by the light electric field.

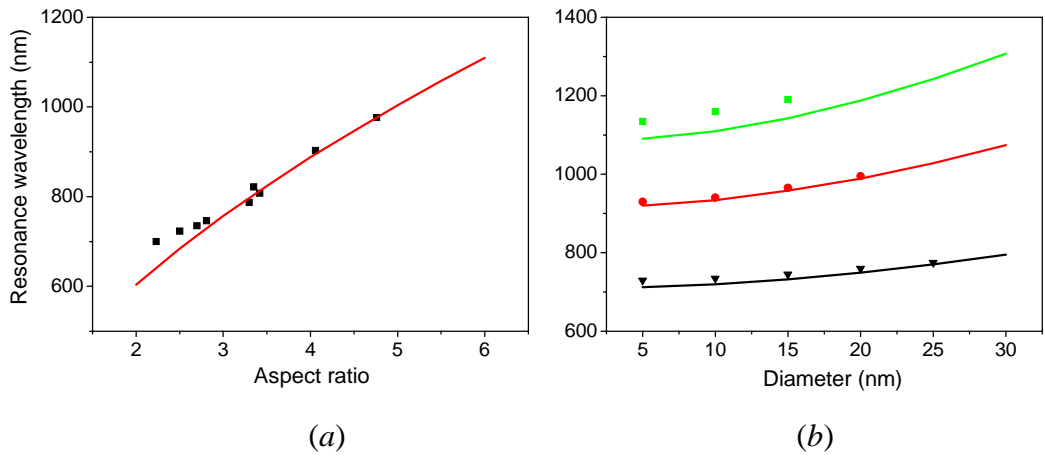

Figure 2: (a) Dependence of resonance wavelength on the aspect ratio. The squares represent the experimental data [10] and the line gives our calculated values (the rod diameter is fixed as 22nm). (b) Dependence of resonance wavelength on the rod radius. The symbols are obtained by numerical calculations [19] and the lines obtained by our calculation. Here the aspect ratio is set as 3, 5 and 7, respectively (from the bottom to the top).

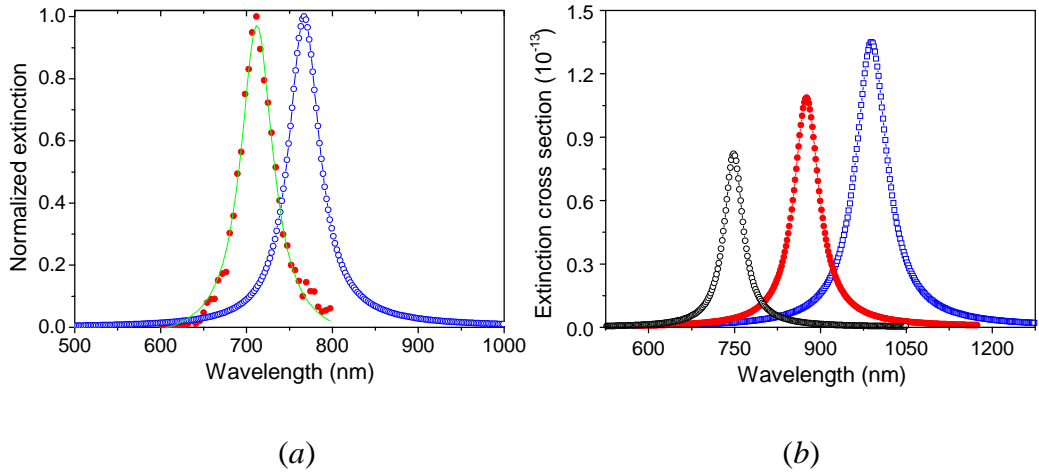

Figure 3: (a) Normalized extinction spectra for a single gold nanorod, which has a length of 52.65nm and radius of 8.1nm. The open and solid circles represent, respectively, the theoretical (by equation (5)) and experimental [11] results. The solid line is a numerical fit of experimental data with the Lorentzian line-shape. (b) Calculated extinction spectra for three gold nanorods with different rod sizes, where the rod radius is fixed as 10nm and the length is set as 60, 80, and 100nm, respectively (from the left to the right).